\documentclass[global,referee]{svjour}

\usepackage{graphics}
\journalname{apb}

\begin{document} 
 
\title{A Femtosecond Neutron Source}
\author{Andrea Macchi}
\institute{polyLAB, CNR-INFM, Dipartimento di Fisica ``E. Fermi'',
Universit\`a di Pisa, Largo B.~Pontecorvo 3, 56127 Pisa, Italy.
\email{macchi@df.unipi.it} (Fax: +390502214333)}

\date{\today}

\maketitle 
 
\begin{abstract} 
The possibility to use the ultrashort ion bunches produced by circularly 
polarized laser pulses to drive a source of fusion neutrons with sub-optical 
cycle duration is discussed. A two-side irradiation of a deuterated thin foil 
target produces two countermoving ion bunches, whose collision produces an 
ultrashort neutron burst. Using particle-in-cell simulations and analytical 
modeling, it is evaluated that, for intensities of a few 
$10^{19}~\mbox{W cm}^{-2}$, more than $10^3$ neutrons per Joule may be 
produced within a time shorter than one femtosecond. Another scheme based on 
a layered deuterium-tritium target is outlined.
\end{abstract}

\section{Introduction}
With the advent of short-pulse laser systems yielding multi-terawatt power,
laser-driven nuclear physics has emerged as a very active area of research
\cite{ledingham} with applications such as radioactive isotope production
and nuclear transmutation of elements. In particular, the emission of neutrons
from fusion reactions has been observed in several experiments with different
pulse parameters and target types, including solid targets 
\cite{solid},
heavy water droplets 
\cite{droplets},
deuterium clusters 
\cite{clusters,madison}, 
and underdense plasmas or gas jets 
\cite{underdense}.
In these experiments, 
the number of neutrons produced per Joule of the laser pulse energy
is usually in the $10^4-10^5~\mbox{J}^{-1}$ range for terawatt, 
femtosecond lasers, and higher for large petawatt, 
picosecond lasers (see table~I of Ref.\cite{madison} for a partial summary).
While the large neutron fluxes produced by petawatt pulses are of importance
for material damage studies relevant to thermonuclear fusion research
\cite{perkins}, neutron bursts produced by  ``table-top terawatt'' 
(T$^3$) lasers at high repetition rate may provide compact, pulsed
neutron sources for radiography and other applications. 

The interpretation of the above experiments suggests that, 
as a general issue, the route to neutron production starts from the
heating of target electrons up to high energies; in turn, the electron 
currents produce space-charge fields driving ion acceleration up to MeV 
energies (via various mechanisms such as sheath acceleration in solid targets, 
or Coulomb explosion in clusters and underdense plasma channels);
finally, collisions between ions lead to fusion reactions and
neutron emission. The duration of the latter
could not be measured so far in experiments; however, 
based on the physical picture which is inferred from the experiments, 
we expect the neutron pulse duration not to be shorter than the laser pulse
duration, i.e., to be typically in the $0.1\div 1.0~\mbox{ps}$ range. 

In this paper, we study a new approach to neutron
production by ultrashort laser pulses, which could provide a source 
of fusion neutrons with sub-optical cycle duration.
This represents another possible route to the ultrafast control and imaging 
of nuclear reactions by superintense fields
\cite{nucontrol} and may be used to study phenomena such as
nuclear spin-mixing oscillations
whose period was estimated to be $\sim 1~\mbox{fs}$ 
\cite{pachucki}. 

Our approach is based on the prediction that
the irradiation of a solid targets with circularly polarized, intense 
short pulses leads to the prompt acceleration
of high-density, short duration
ion bunches \cite{macchi}.
Such bunches may be used efficiently as projectiles for beam fusion,
since for deuterium the energy in the center-of-mass system 
may approach the Gamow value 
(${\cal E}_G\approx 1~\mbox{MeV}m_r/m_p$, where $m_r$ is the reduced mass) 
for which the cross sections
of deuterium-deuterium (D-D) or deuterium-tritium (D-T)
reactions have a maximum.
With an appropriate target scheme, fusion reactions and related
neutron emission may last just for a time of the order of the
ion bunch duration, which may be less than one optical cycle.

\section{The colliding bunches scheme}

To find a scheme based on the 
$\mbox{D}+\mbox{D}\rightarrow {}^3\mbox{He}+\mbox{n}~(2.45~\mbox{MeV})$ 
reaction aiming at the shortest achievable duration,
we consider a symmetrical, double-sided irradiation of a thin foil target.
In this scheme, two colliding ion bunches are generated.
Thus, if the two bunches are properly timed, for a given value of the
laser intensity the energy in the center-of-mass is maximized while 
the duration of neutron emission is minimized.
This experimental geometry is similar to the one 
of Ref.\cite{shen}, where  
a ``laser-confined'' thermonuclear fusion approach was proposed.
In that scheme, a thin deuterium-tritium foil is compressed, heated and
confined by double-side irradiation using relatively long pulses.
In our scheme, to produce two single bunches
the laser pulse duration must be much shorter (a few cycles) 
than in Ref.\cite{shen}. Our approach is based on 
a non-thermal \emph{beam fusion} concept
with emphasis on the ultrashort duration of the neutron emission, 
and without concerns of target stability.

We now discuss the target and pulse requirements based on the
features of ion bunch generation studied in Ref.\cite{macchi}.
The bunch ion velocity spectrum extends up to a maximum velocity 
$v_m$, given by
\begin{equation}\label{eq:vscaling} 
\frac{v_m}{c} = 2\sqrt{\frac{Z}{A}\frac{m_e}{m_p}\frac{n_c}{n_e}}a_L 
\simeq 0.047 \sqrt{\frac{Z}{A}\frac{n_c}{n_e}}a_L \, ,
\end{equation}
$n_e$ is the initial electron density,
$n_c=m_e\omega_L^2/4\pi e^2$ is the cut-off density for the laser
frequency $\omega_L$, and 
$a_L=0.85\sqrt{I\lambda^2/10^{18}~\mbox{W cm}^{-2}\mu\mbox{m}^2}$ is the 
dimensionless laser amplitude, being $I$ 
and $\lambda=2\pi c/\omega_L$ the intensity and the wavelength of the
laser pulse, respectively. 
At a given value of $a_L$, the lower $n_e$ the higher $v_m$.
Collisions in the center-of-mass system at 
the Gamow energy for the D-D reaction will begin when
$2v_m=v_G=\sqrt{2{\cal E}_G/m_p}\simeq 0.0458c$.
The ion bunch acceleration occurs over a typical time 
$\tau_i \simeq 7T_L({A}/{Z})^{1/2}{a_L}^{-1}$ where $T_L$ is the 
duration of a laser cycle.
Thus, ultraintense few-cycle pulses, which are now at the frontier of
current research \cite{krausz}, are best suited for ion bunch 
acceleration. 

The number of accelerated ions per unit surface is 
$\simeq n_{i0} l_s$, where $n_{i0}$ is the background ion density
and $l_s$ is the evanescence length 
of the ponderomotive force inside the plasma.
At the end of the acceleration stage ($t=\tau_i$),
a sharp peak of the ion density is formed 
at a distance $\simeq l_s$ from the original target surface;
the peak density is $n_b=\kappa n_0$, 
where $\kappa \sim 10$ is found from simulations,
and the bunch width is thus $l_b \simeq l_s/\kappa$.
One expects $l_s \simeq c/\omega_p=(\lambda/2\pi)\sqrt{n_c/n_e}$,
where $\omega_p=\sqrt{4\pi n_e e^2/m_e}$ is the plasma frequency.
Thus, $l_b \ll \lambda$ holds when $n_e \gg n_c$,
and the bunch duration $\tau_b \simeq l_b/v_m$ and can be less 
than $T_L=\lambda/c$.

In the colliding bunches scheme the duration of the neutron burst,
will be of the order of $\tau_b/2$, i.e. in the sub-femtosecond
regime. The optimal target thickness is close to the 
value $\ell\simeq 2l_s$, in order
to let the two bunches collide at the end of the acceleration stage.
Since a very thin foil target is needed, 
deuterated plastic, e.g. $(\mbox{CD}_2)_n$, looks more
suitable than pure (cryogenic) deuterium as a target material; 
however, the much higher value of
$n_e$ requires higher pulse intensities.
In the following, both pure solid D 
($n_i=n_e \simeq n_0=6\times 10^{22}~\mbox{cm}^{-3}$ i.e. $n_e=40n_c$ 
for $\lambda=0.8~\mu\mbox{m}$)
and ``plastic'' targets (with same $n_i=n_0$ but $n_e=250n_c$)
are discussed. 

\section{Simulation results}

We now analyze the double-sided irradiation of a thin foil 
using a particle-in-cell (PIC) simulation. 
As shown in Ref.\cite{macchi} a one-dimensional (1D) model is adequate
to keep the essential features of ion bunch generation. This allows us
to use a very fine spatial and temporal resolution, which is actually
necessary to resolve the bunch formation properly, since the latter is 
characterized by sharp density gradients and very short formation times.
In our simulations we use 2000 grid cells per wavelength and 2000 particles
per cell at the initial time. 
In addition, to compute the neutron yield from the simulation data 
(as shown below) properly, the ion distribution function has to be 
reconstructed over a phase space grid once every a few timesteps.

\begin{figure}
\resizebox{0.75\textwidth}{!}{\includegraphics{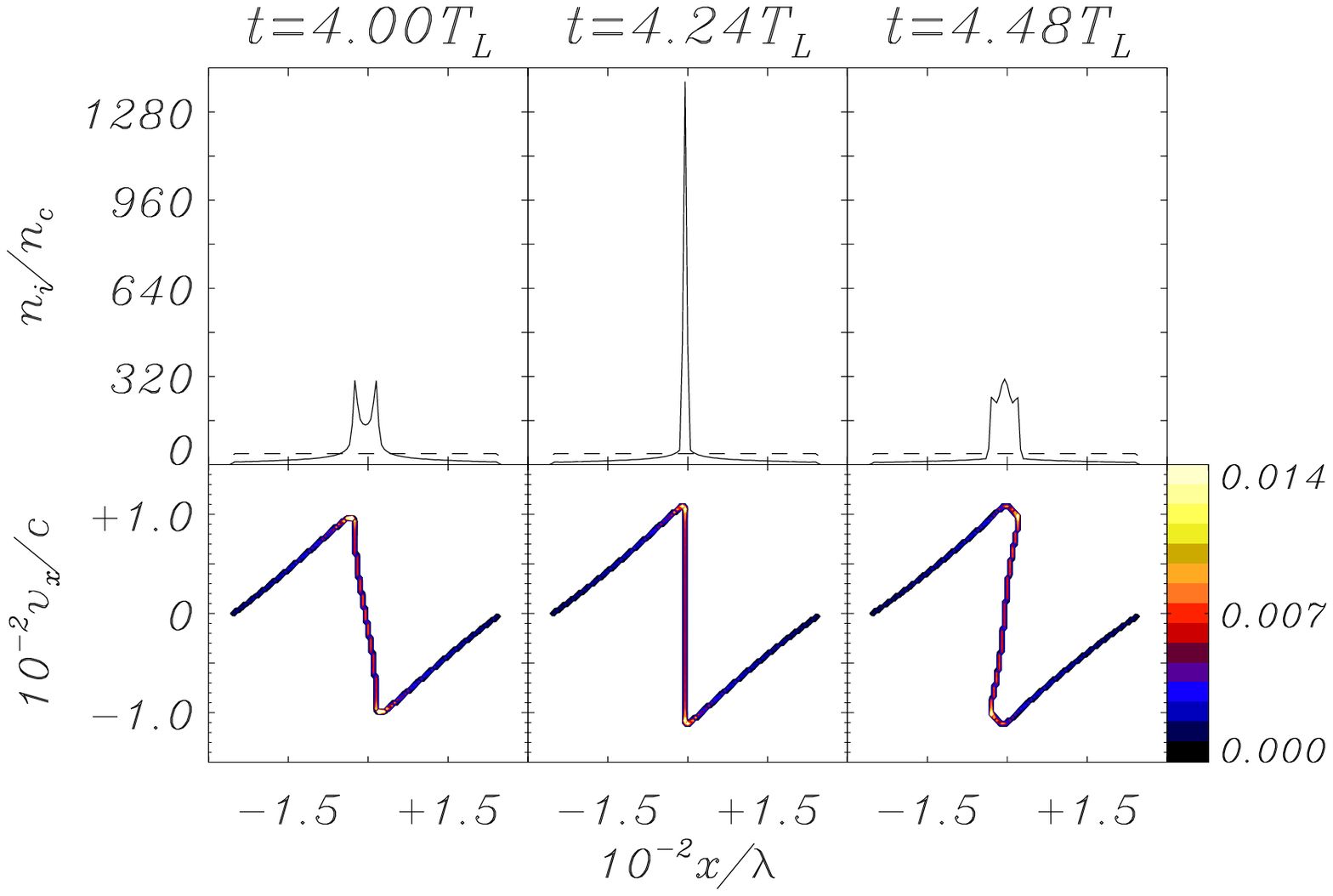}}
\caption{(Color online.) The ion density $n_i$ (top row, solid line) 
and the phase space distribution $f(x,v_x)$ (bottom row, arbitrary units) 
at different times (labels)  
from an 1D PIC simulation of two-side irradiation of a thin foil. 
The initial density profile is also shown (dashed line).
Run parameters are $a_L=2.5$, $n_{0}/n_c=40$, 
and $\ell=0.05\lambda$. 
} \label{fig:pic}  
\end{figure}

Figure~\ref{fig:pic} shows results for a ``D'' foil  
simulation. The front and rear laser pulses both 
have peak amplitude $a_L=2.5$ and duration $\tau_L=5T_L$ (FWHM), 
corresponding to an intensity of 
$1.3\times 10^{19}~\mbox{W cm}^{-2}$, 
and a duration of $13~\mbox{fs}$ if $\lambda=0.8~\mu\mbox{m}$.
The foil target has initial density $n_e=n_{i0}=40n_c$ and 
thickness $\ell=0.05\lambda$.
The ion density reaches a peak value close to $1390n_c \simeq 35n_{i0}$,
i.e. $\kappa \simeq 17$,
with rise and fall times of $\approx 0.1T_L$. The maximum ion 
velocity $v_m \simeq 10^{-2}c$. The results of a ``CD$_2$''
simulation ($n_e/n_c=250$, $a_L=8$, $\ell=0.02\lambda$ 
and all the other parameters equal to the case of Fig.\ref{fig:pic})
are qualitatively similar:
$\kappa\simeq 10$ and $v_m\simeq 0.01c$ are found.

We now compute the neutron burst intensity and
duration from the simulation data.
The reactivity for a fusion reaction \cite{atzeni} is given by 
\begin{equation}
\langle\sigma v\rangle=\int v\sigma(v)g(v)dv \, ,
\label{eq:reactivity}
\end{equation}
where $\sigma(v)$ and $g(v)$ are the cross section and the distribution
function (normalized to unity), respectively, in terms of the relative
velocity $v=|{\bf v}_1-{\bf v}_2|$, being ${\bf v}_{1,2}$ the velocities
of the two ion species. The number of fusion reactions
per unit volume and time is given by
$R=n_1 n_2\langle\sigma v\rangle/(1+\delta_{12})$ \cite{atzeni}, 
being $n_{1,2}$ the number densities
of the two ion species. For ions of the same species, as in the D--D
reaction, $n_1=n_2=n$. 
A convenient parameterization of the cross section
is given by \cite{atzeni}
\begin{equation}
\sigma=\frac{S({\cal E})}{{\cal E}}e^{-\sqrt{{\cal E}_G/{\cal E}}}\, ,
\label{eq:sigma}
\end{equation}
where ${\cal E}=m_r v^2/2$ is the center-of-mass energy, 
and $S$ is the astrophysical factor that is a slowly varying
function of ${\cal E}$ and hence will be taken as a constant 
$S_0=5.4\times 10^{-23}~\mbox{keV~cm}^2$ for D-D.

From the PIC simulation data we obtain $g(v)$ at any grid point 
by computing the ion velocity distribution $f(v_i)$
on a phase space grid, from which
$g(v)$ is obtained by convolution. We thus compute the reactivity
and, from the knowledge of the ion density, the number of fusion reactions
per unit density and time at each grid point. Finally, we obtain the 
total rate of fusion events and the overall number of neutrons produced
by integrating $R$ over space. 

\begin{figure}
\resizebox{0.75\textwidth}{!}{\includegraphics{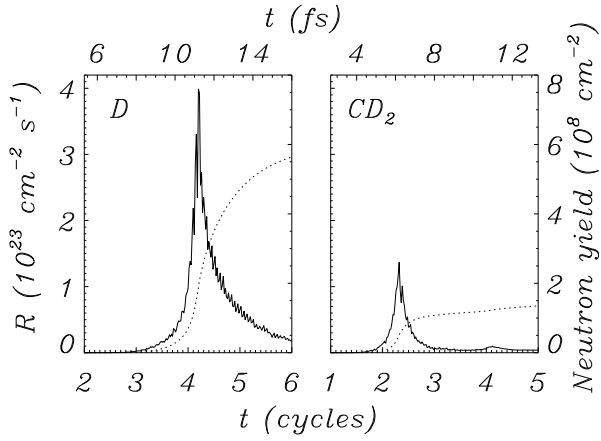}}
\caption{Numerical evaluation of
the rate (solid line) and the total number (dotted line) of neutrons
produced per unit surface for the ``D'' simulation of Fig.\ref{fig:pic}
(left panel) and for the ``CD$_2$'' simulation for which 
$n_e=250n_c$, $a_L=8$.} 
\label{fig:burst}  
\end{figure}

The results are shown in Fig.\ref{fig:burst} for both the ``D'' and 
``CD$_2$'' target cases. The same number density of deuterium
ions has been assumed. 
In both cases, a neutron burst is generated 
with a duration of about $0.7~\mbox{fs}$ (FWHM) and a yield
of $\sim 10^9\mbox{neutrons~cm}^{-2}$.
Since the pulse duration is $\simeq 15~\mbox{fs}$,
the number of neutrons produced per Joule of the 
pulse energy is
$\sim 10^{3}~\mbox{J}^{-1}$ for the D 
case with $a_L=2.5$,  $\sim 10^{3}~\mbox{J}^{-1}$ CD$_2$ case
($a_L=8$);
these numbers are roughly between one and two orders of magnitude lower 
than that inferred from experiments with T$^3$ systems; however, 
for the present scheme the
expected duration is likely to be much shorter, and
the emission rate may be comparable or even higher. 

The relatively long-lasting tail in the fusion rate is due to
the fact that the two bunches, after crossing each other, propagate
in a low-density shelf of ions which originate from the layer 
of charge depletion at the surface
and have been accelerated to lower energies than the bunch ions 
(see Fig.\ref{fig:pic} and Ref.\cite{macchi}). 
At higher intensities the ions in the shelf have
enough energy to sustain a significant rate of fusion reactions during
the propagation of the bunch, leading to a higher total yield but also
to a longer duration (a few femtoseconds) of the neutron emission. 

\section{Analytical scalings}

To check our numerical results, and to provide an approximate scaling of
the neutron yield versus the laser intensity, we 
analytically evaluate the fusion rate in
the colliding-bunches scheme. 
From Ref.\cite{macchi} we infer that it is worth to 
consider for the ion bunch velocity distribution $f(v_i)$ 
either a ``flat-top'' velocity distribution 
$f_{F}=1/v_m$ for $0<v_i<v_m$ and zero elsewhere, 
and a monochromatic distribution 
$f_D=\delta(v_i-v_m)$.
In both cases the relative velocity distribution 
$g(v)$ is easily obtained analytically by convolution.
The averaged reactivity may thus be written as
\begin{eqnarray}
\langle\sigma v\rangle 
=\frac{4S_0}{m_p{v}_G}\int \frac{e^{-v_G/v}}{v/v_G}g(v)dv
\equiv \frac{4S_0}{m_p{v}_G}{\cal M}(\zeta)
 \, ,
\end{eqnarray}
where $\zeta={v_G/v_m}$, 
\begin{eqnarray}
{\cal M}(\zeta)
=\left\{\begin{array}{lr}
2\zeta\left[\mbox{E}_1(\zeta/2)-\mbox{E}_1(\zeta)
     -\mbox{E}_2(\zeta/2)+\mbox{E}_2(\zeta)\right]\, , & (f=f_F)\, , \\ 
(\zeta /2) e^{-\zeta/2} \, ,& (f=f_D)\, ,
\end{array}\right. 
\end{eqnarray}
and $\mbox{E}_n(x)=\int_{1}^{\infty}(e^{-xt}/t^n)dt$ is the
exponential integral function of order $n$.

\begin{figure}
\resizebox{0.75\textwidth}{!}{\includegraphics{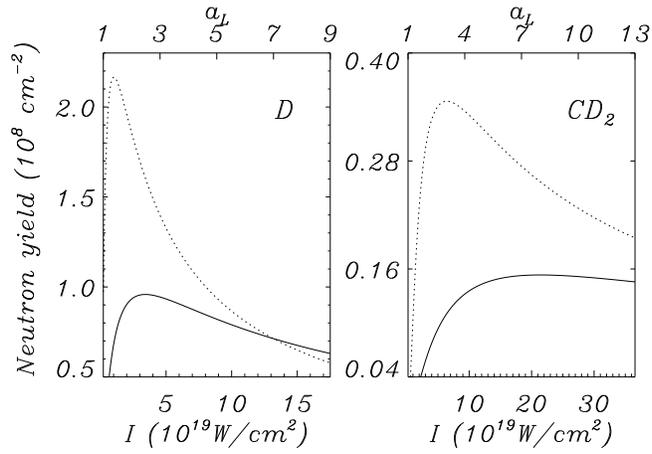}}
\caption{Analytical scaling of the number of neutrons (per unit surface)
produced \emph{during the ultrashort burst} as a function
of laser intensity for a $\lambda=0.8~\mu\mbox{m}$ pulse,
for a ``D'' target with $n_e=40n_c$ and a ``CD$_2$'' target with
$n_e=250n_c$. Solid and dotted curves correspond to the choice
of ``flat-top'' and monochromatic distribution functions, respectively.} 
\label{fig:yield}  
\end{figure}

The rate of neutrons produced per unit volume is given by 
$R=(n_i^2/2)\langle\sigma v\rangle$, 
where $n_i=\kappa n_0$.
To obtain the number $N$ of neutrons produced per unit surface
we multiply $R$ by the spatial extension of the 
neutron burst, $l_b \simeq l_s/\kappa$,
and by the burst duration $\simeq \tau_b/2 \simeq l_s/2\kappa v_m$.
Hence, the total 
neutron yield does \emph{not} depend on the compression ratio 
$\kappa$, which only affects the bunch width and the 
burst duration. We obtain
\begin{equation}
N \simeq \frac{n_0^2}{2}\frac{l_s^2}{2v_G}\frac{v_G}{v_m}
\frac{4S_0}{m_p{v}_G}
{\cal M}(\zeta)
   \equiv N_0 \zeta{\cal M}(\zeta) \, .
\label{eq:yield}
\end{equation}
The dependence of $N$ on the laser intensity 
is shown in Fig.\ref{fig:yield} for both choices of $f(v)$.
Posing $l_s \simeq c/\omega_p$, we obtain
$N_0 \simeq 2 \times 10^{8}$ for ``D'', 
$N_0 \simeq 3 \times 10^{7}$ for ``CD$_2$'' targets.
These values underestimate the neutron yield observed in simulations;
this is likely to be due to the fact that the effective screening length
$l_s$ is actually larger than $c/\omega_p$ and/or the active volume 
for fusion reactions is wider than the bunch, since, as observed above,
fusion reactions occur also in the low-density shelf.
This explanation is also supported by noting that 
the burst duration is also underestimated
when taking $l_s \simeq c/\omega_p$:
for ``D'' targets,
$c/(2\omega_p\kappa v_m)\simeq 0.2~\mbox{fs}$ while a duration of 
$0.8~\mbox{fs}$ is observed in Fig.\ref{fig:burst}.

\section{Single bunch scheme}

Using ultrashort ion bunches, it may be also possible to obtain 
a fetmosecond source of neutrons or other fusion products 
from heteronuclear reactions
using a single short pulse impinging on a layered target. As an example
we consider neutron production
in a target with a thin surface layer of deuterium for ion 
acceleration and an inner tritium layer as an immobile target,
using the 
$\mbox{D}+\mbox{T}\rightarrow \alpha+\mbox{n}~(14~\mbox{MeV})$ reaction.
In the relevant energy range\footnote{The D-T cross section 
has a broad maximum
around $64~\mbox{keV}$; however, this low-energy range 
(where $S({\cal E})$ varies strongly with energy)
is not considered here,
since the laser-plasma interaction regime becomes very collisional
and must be further investigated.}
the astrophysical factor for the D-T reaction is
$S_0=1.2\times 10^{-20}~\mbox{keV~cm}^2$, almost 200 times larger than
for D-D. This effect compensates the loss in the center-of-mass energy 
with respect to the colliding bunches scheme at the same laser intensity.
If the deuterium layer has a thickness $\simeq l_s$, and the tritium
layer is not thicker than the ion bunch, the neutron burst duration
might be limited to $\sim l_b/v_m$. An analytical estimate, analogous
to that leading to Eq.(\ref{eq:yield}), leads to the following result 
\begin{eqnarray}
N \simeq n_0^2\sqrt{\frac{10}{3}}\left(\frac{S_0}{m_pv_G}\right)_{DT} 
\frac{l_s^2}{\kappa v_m}{\cal A}({\zeta}) 
\simeq 1.3 \times 10^{11}~\mbox{cm}^{-2}\kappa^{-1}\zeta{\cal A}({\zeta})\, ,
\end{eqnarray}
where ${\cal A}({\zeta})=a\zeta \mbox{E}_1(a\zeta)$ or $a\zeta e^{-a\zeta}$
for a flat-top or monochromatic ion distribution, respectively,
$a=\sqrt{125/24}$, and $n_D \simeq n_T \simeq n_0$ has been assumed.
If the tritium layer $l_T$ is thicker than $l_b$, both the yield and the
duration increase by a factor $\sim l_T/l_b$.
The function $\zeta{\cal A}({\zeta})$ has a maximum in the range of 
intensities $10^{20} \div 10^{21}~\mbox{W cm}^{-2}$.

\section{Discussion and conclusions}

We briefly discuss the experimental feasibility
of the proposed scheme.
The required laser pulse duration and intensity are within present-day, or
at least near-term capabilities of table-top systems.
Quenching of prepulse, pulse synchronization and need of circular polarization 
as well as thin foil target manifacturing are likely to be demanding tasks
but not out of reach.   
The measurement of the duration of a neutron burst with sub-femtosecond 
resolution is a challenging issue. 
We argue that it may be possible to perform
an indirect measurement based on the interaction of the 
neutrons with a secondary target, where charged particles are
produced; this process may be resolved in time
by using attosecond spectroscopy techniques
\cite{attoseconds}.

In conclusion, we investigated the production of fusion neutrons
using ion bunches accelerated in solid deuterated targets 
by circularly polarized laser pulses. Our results, 
in connection with the continuous progress 
in producing ultrashort, superintense laser pulses at a high repetition rate, 
may open a perspective for a neutron source with
sub-optical cycle duration and sufficient brightness for applications.

\textsc{acknowledgments} The author is grateful to 
S.~Atzeni, D.~Bauer, F.~Ceccherini, F.~Cornolti, and F.~Pegoraro 
for critical reading and enlightening discussions. 
Support from the INFM supercomputing initiative is also acknowledged.

\newpage

\end{document}